\documentstyle[12pt]{article}

\begin{document}

\begin{center}

ISOVECTOR VIBRATIONS IN NUCLEAR MATTER AT FINITE TEMPERATURE

\vspace{0.5cm}

M. Di Toro$^{1)}$, V.M. Kolomietz$^{2,3)}$, A.B. Larionov$^{1,4)}$

\end{center}

\vspace{0.5cm}

\noindent
$^{1)}$Dipartimento di Fisica and Istituto Nazionale
di Fisica Nucleare, Laboratorio Nazionale del Sud,
95123 Catania, Italy

\noindent
$^{2)}$Institute for Nuclear Research, Prosp. Nauki 47,
252028 Kiev, Ukraine

\noindent
$^{3)}$Cyclotron Institute, Texas A\&M University,
College Station TX 77843-3366, USA

\noindent
$^{4)}$Kurchatov Institute, Moscow 123182, Russia

\vspace{0.5cm}

\begin{abstract}
We consider the propagation and damping of isovector 
excitations in heated nuclear matter within the Landau 
Fermi-liquid theory. Results obtained for nuclear matter are applied 
to calculate the Giant Dipole Resonance (GDR) at finite 
temperature in heavy spherical nuclei within Steinwedel 
and Jensen model. 

The centroid energy of the GDR slightly decreases with increasing 
temperature and the width increases as $T^2$ for temperatures
$T < 5$ MeV in agreement with recent experimental data for
GDR in $^{208}$Pb and $^{120}$Sn.

The validity of the method for other Fermi fluids is finally
suggested.
\end{abstract}

\vspace{0.5cm}


\vspace{0.5cm}

In recent years the GDR built on highly excited states is in the center 
of many experimental and theoretical studies (c.f. \cite{Pierr} and 
references therein). In this context, one of the most important open 
problems is the behaviour of the GDR width in {\it nonrotating} nuclei 
as a function of temperature. There are two essentially different 
theoretical approaches to this problem. The first one \cite{OBB96}
explains the temperature increasing of the width as an effect of the
adiabatic coupling of the GDR to thermal shape deformations. In the
second approach \cite{SBD91,Vir97,DS97} the thermal contribution to the 
damping width arises from an increasing nucleon-nucleon collision 
rate ($2p2h$ excitations) 
 plus a Landau spreading due to thermally allowed 
$ph$ transitions \cite{Vin92,BV,BVA,KLD97}. 
 
In the present work, following the ideology of the second approach, 
we consider isovector volume vibrations in spin-isospin symmetrical 
nuclear matter at finite temperature. A similar problem was considered
in Refs. \cite{BV,BVA} within the RPA method. However the Landau
damping mechanism of the dissipation of a propagating mode
due to thermal smearing of Fermi distribution is too weak to be responsible 
for the fast increase of the observed GDR width with temperature 
\cite{BV,BVA,Klaus}. 
This problem can be solved by taking into account the two-body 
dissipation through the collision integral of the Landau-Vlasov equation 
\cite{SBD91}. The use of a quantum kinetic equation leads to
memory effects in the collision term in order to include off 
energy-shell contributions \cite{AK}.
Moreover it was shown in Refs. \cite{Ayik,KPS96}, that memory 
effects are essentially increasing the widths of 
multipole resonances at small temperatures. In this Rapid Communication we 
calculate the isovector strength function of nuclear matter taking into 
account both thermal Landau damping and two-body collisional dissipation,
including the quantum memory contribution.
        
The isovector response of uniform nuclear matter is described by the
linearized Landau-Vlasov equation with a collision term treated 
in the relaxation
time approximation \cite{Vir97,KLD97,KPS96}
\begin{equation}
{\partial \over \partial t}\delta f
+ {\bf v}\cdot\nabla_r \delta f
- \nabla_r(\delta U
+ 2 \delta V)\cdot\nabla_p f_{\rm eq}
= -{1 \over \tau}\delta f|_{l\geq1}~,                   \label{LVe}
\end{equation}
where $\delta f \equiv \delta f_n - \delta f_p$
and $\delta U \equiv \delta U_n - \delta U_p$ are
differences between neutron and proton distribution functions (d.f.)
and mean fields respectively,
$\delta V \equiv \delta V_n - \delta V_p$ is external field
($\delta V_q = \tau_q\delta V,~~\tau_n=+1,~~\tau_p=-1$)
\cite{BV}, $f_{eq}(\epsilon_p = p^2/2m$) is the equilibrium finite 
temperature Fermi distribution, and the notation $l\geq1$ means
that the perturbation of the d.f. $\delta f|_{l\geq1}$ in collision 
integral includes only Fermi surface distortions with multipolarity 
$l\geq1$ in order to conserve the particle number in collision processes
\cite{AK}. The inclusion of the $l=1$ harmonic in the collision integral of 
Eq. (\ref{LVe}), at variance with the isoscalar case \cite{KLD97}, 
is due to nonconservation of the isovector current, i.e. due to a collisional
friction force between counterstreaming neutron and proton flows.

The dynamical component of the isovector mean field $\delta U$ can be
expressed in terms of the isovector Landau parameter $F_0'$ : 
\begin{equation}
\delta U = {F_0' \over N(T)}\delta\rho~,              \label{dU}
\end{equation}
where
\begin{equation}
\delta\rho({\bf r};t)
= \int\,{g d{\bf p} \over (2\pi\hbar)^3}\,
\delta f({\bf r},{\bf p};t)~,                         \label{drho}
\end{equation}
is the density perturbation, $g=2$ is the spin degeneracy factor and
\begin{equation}
N(T) = \int\,{g d{\bf p} \over (2\pi\hbar)^3 }\,
\left(
- { \partial f_{\rm eq}(\epsilon_p) \over
    \partial {\epsilon_p}                   }
\right)                                              \label{NT}
\end{equation}
is the thermally averaged density of states, 
$N(0) = g p_F m / 2\pi^2\hbar^3$, where we put for simplicity 
$m^* = m = 938$ MeV. 

For an  external field
$\delta V \propto \exp(i{\bf kr} - i\omega t)$, 
periodic in space and time, the isovector collective
response function \cite{BV} can be derived from Eq. (\ref{LVe}):
\begin{equation}
\chi^{coll}(\omega,{\bf k})
= -{\delta\rho \over \delta V}
= { 2 N(T)\chi_T^\tau \over 1 + F_0'\,\chi_T^{\tau} }~,  \label{chicoll}
\end{equation}
where $\chi_T^\tau$ is the intrinsic response function \cite{Hofm92,Yam94}.
The explicit form of the function $\chi_T^\tau(\omega,{\bf k})$ is
(details of derivation in Ref. \cite{KLD97}):
\begin{equation}
\chi_T^\tau(s) =
- {N(0) \over m p_F N(T)} \int\limits_0^\infty dp \,\,
{p^2 s \,\chi(\overline{p}s/p) \over
s^\prime + i s^{\prime\prime} \chi(\overline{p}s/p)}\,
{\partial f_{\rm eq}(\epsilon_p) \over \partial \epsilon_p}~,
                                                           \label{chitau}
\end{equation}
where
\begin{eqnarray}
& & \overline{p}
= p_F\left({\overline{\epsilon} \over \epsilon_F}\right)^{1/2}~,
							\label{pav} \\
& & \overline{\epsilon}
= {5 \over 3\rho_{\rm eq}} \int\,{g d{\bf p} \over (2\pi\hbar)^3}\,
\epsilon_p f_{\rm eq}(\epsilon_p)~,                     \label{ekin} \\
& & \rho_{\rm eq}
= \int\,{g d{\bf p} \over (2\pi\hbar)^3}\,f_{\rm eq}(\epsilon_p)
                                                         \label{dens}
\end{eqnarray}
are quasiparticle average momentum, average kinetic energy  
( respectively normalized
at $T=0$ on $p_F$ and $\epsilon_F$) and density, with the complex
variable

\begin{equation}
s = s^\prime + i s^{\prime\prime},\,\,\,\,\,\,
s^{\prime\prime} = {m \over \tau \overline{p} k},\,\,\,\,\,\,
s^\prime = {\omega m \over \overline{p} k}~,                \label{sprime}
\end{equation}
$\chi(z)$ is a Legendre function of the second kind
\begin{equation}
\chi(z) = {1\over 2} \int\limits_{-1}^1 d\mu \,{\mu \over
{\mu - z}}~.
                                                             \label{chi}
\end{equation}
Eq. (\ref{chitau}) for the intrinsic response function of 
isovector vibrations
has only a minor difference with isoscalar case. Namely,
to recover the isoscalar response function given by Eq. (30) in Ref. 
\cite{KLD97}, one should change 
$i s^{\prime\prime} 
\rightarrow i s^{\prime\prime}(1 + 3s^{\prime}s\overline{p}^2/p^2)$
in the denominator of Eq. (\ref{chitau}). This difference is just due to 
inclusion of the damping of the $l=1$ harmonic in the isovector channel.
We note, that the r.h.s. of Eq. (30) in Ref. \cite{KLD97} has an error: 
it should be multiplied by a minus sign.

For a given momentum transfer $k$, the strength function per unit
volume is:
\begin{equation}
S_k(\omega) = {1 \over \pi}\mbox{Im}(\chi^{coll})
= { 2 N(T)\,\mbox{Im}(\chi_T^\tau)/\pi \over
(1 + F_0'\,\mbox{Re}(\chi_T^\tau))^2
+ (F_0'\,\mbox{Im}(\chi_T^\tau))^2 }~.
						 \label{strength}
\end{equation}
The strength function satisfies the following energy weighted sum rule \\
(EWSR) \cite{BV,BVA}:
\begin{equation}
\int\limits_0^\infty\,d\omega \omega S_k(\omega)
= {k^2 \over 2m}\rho_0~,                      \label{EWSR}
\end{equation}
where $\rho_0 = 0.16$ fm$^{-3}$ is the nuclear saturation density. 
 
Collective modes are given by poles of the response function
(\ref{chicoll}):
\begin{equation}
1 + F_0'\,\chi_T^\tau(s) = 0~.                        \label{drel}
\end{equation}
By solving Eq. (\ref{drel})  we obtain the complex
frequency~:
\begin{equation}
\omega = \omega_R + i\omega_I = k{\overline{p} \over m}
                                (s - is^{\prime\prime})~.
  \label{omega}
\end{equation}

The application of the formalism discussed above to finite nuclei
is based on the Steinwedel-Jensen (SJ) picture \cite{BV,BVA,RS} 
which describes the GDR in heavy nuclei as a volume polarization mode
conserving the total density $\rho = \rho_n + \rho_p$. According
to this model, we choose the wavenumber of the normal mode as
 $k=\pi/2R$, where $R$ is the radius of 
a nucleus. Inside the nucleus, the unperturbed distribution of nucleons 
is supposed to be uniform. The SJ model gives a good overall reproduction
of the ground state GDR energies for heavy spherical nuclei \cite{Berman}.

We have to remark that the calculation of GDR widths in finite nuclei
is a much more difficult problem. It can be only partially solved within 
our nuclear matter approach, since shell effects
and the escape width due to particle emission are 
not taken into account in the present investigation. However in
the temperature region $T = 1 \div 3$ MeV the shell effects 
start to smear out and the escape width is still very small.
Thus we expect our calculations to be quite reliable in 
this temperature region. 

Eq. (\ref{LVe}) contains two free parameters: the isovector Landau
parameter $F_0'$ and the two-body relaxation time $\tau$.

The Landau parameter $F_0'$ at zero temperature can be
expressed as a function of the symmetry energy coefficient $\beta$
in the Weizs\"acker mass formula as follows \cite{Mig}:
\begin{equation}
F_0'(T=0) = {3\beta \over \epsilon_F} - 1~.               \label{Migform}
\end{equation}
For the standard value of $\beta=28$ MeV we have $F_0'(0)=1.33$.
The value of the $F_0'$ decreases with temperature due to the decrease
of the level density $N(T)$:
\begin{equation}
F_0'(T) \simeq F_0'(0)\left[ 1 - 
{\pi^2 \over 12}\left({T \over \epsilon_F}\right)^2 \right]~.  \label{Ftemp}
\end{equation} 
However the coupling constant $F_0'(T)/N(T)$ is independent on temperature.

The two-body relaxation time $\tau$ includes temperature and memory 
effects~: 
\begin{equation}
\tau = { \hbar\alpha^{(-)} \over T^2
+ (\hbar\omega_R/2\pi)^2       }~.
                               	                    \label{tau}
\end{equation}
The dependence of the relaxation time on the frequency $\omega_R$ arises 
from memory effects and corresponds to the Landau prescription \cite{AK}.
The coefficient $\alpha^{(-)}$ depends on nucleon-nucleon scattering cross 
sections. We have calculated this coefficient using energy and angular
dependent differential cross sections of $pp$ and $np$ scattering derived
from Bonn A potential both with and without in-medium corrections
\cite{LiMach,Li} (see Appendix). Results are~: $\alpha^{(-)} = 2.3~(5.5)$
MeV in the case of vacuum (in-medium reduced) cross sections.

Fig. 1 shows the photoabsorption cross section by a thermally excited
nucleus $^{208}$Pb, which can be expressed in terms of the strength 
function (\ref{strength}) as follows:
\begin{equation}
\sigma_{abs}(\hbar\omega) = 
{4\pi^2e^2 \over \hbar c k^2 \rho_0} {NZ \over A}
\hbar\omega S_k(\omega)~.                              \label{sigabs}
\end{equation}
Expression (\ref{sigabs}) is obtained from comparison of the EWSR                              
(\ref{EWSR}) and the dipole sum rule \cite{RS}
\begin{equation}
\int\limits_0^\infty\,dE \sigma_{abs}(E) =
{2\pi^2e^2\hbar \over mc} {NZ \over A}~.                 \label{TRK}
\end{equation}
As the temperature is growing, the centroid energy of the GDR
(i.e. the peak of the photoabsorption cross section) is slightly 
shifting to the left and the width is increasing. At variance with
pure mean field predictions \cite{BV,BVA,KLD97}, we see a shift to low
frequencies due to the collision integral in Eq. (\ref{LVe})
(c.f. Refs. \cite{CaiD89,MKHS95,KKS96}).
Indeed at larger temperatures an increasing two-body dissipation should 
reduce the frequency of the collective motion, in close analogy with a 
classical oscillator with a friction force. 

In Fig. 2a (solid lines) we report the temperature dependence of the 
full width half maximum (FWHM) and the centroid energy $E_{GDR}$ as 
functions of temperature, calculated from the photoabsorption cross section 
(\ref{sigabs}) for $^{208}$Pb. 
In parallel, in Fig. 3  (solid lines) the real and imaginary parts of 
the pole of the response function (\ref{chicoll}) are shown.
As far as the collective mode is underdamped, i.e. 
$|\mbox{Im}(\omega)|/\mbox{Re}(\omega) \ll 1$, an approximate relation 
\begin{equation}
\mbox{FWHM} = 2\,|\mbox{Im}(\omega)|                 \label{approx}    
\end{equation}
has to be fulfilled \cite{KPS96}. We see from comparison of Figs. 2a
and 3, that the condition (\ref{approx}) is really satisfied,
corresponding to a Breit-Wigner-like shape of the photoabsorption
strength (see Fig. 1). However, at higher temperatures $T > 4$ MeV 
the photoabsorption strength becomes much more closer to the Lorentzian
shape \cite{Gervais}.

At low temperatures, one can neglect the temperature spreading of the
equilibrium Fermi distribution substituting  
$\delta( \epsilon_F - \epsilon_p )$ instead of \\ 
$(-\partial f_{eq}(\epsilon_p)/\partial \epsilon_p)$ into Eq. (\ref{chitau}).
Thus we obtain the following approximate low temperature expression 
for the intrinsic response function~:
\begin{equation}
\chi_T^\tau(s) \simeq { s \chi(s) \over s^\prime + 
                        i s^{\prime\prime} \chi(s) }~,      \label{chiapr}
\end{equation}
where variables $s,~s^{\prime\prime}$ and $s^\prime$ are defined 
by Eqs.(\ref{sprime}) with the change $\overline{p} \rightarrow p_F$.
We remark that Eq.(\ref{chiapr}) when applied to the case of an electron gas
$(F_0~=~N(T)~4 \pi e^2/k^2)$ gives just the longitudinal
dielectric function
\begin{equation}
\epsilon({\bf k},\omega)~=~1 + F_0~\chi_T^\tau
\end{equation}
obtained by Mermin \cite{Mer70}. This is to stress the more
general framework of the results presented here \cite{Klaus1}.

In the rare collision regime $(\omega_R\tau \gg 1)$, an approximate 
solution of the dispersion relation (\ref{drel}) with the intrinsic response 
function Eq.(\ref{chiapr}) can be found analytically (c.f. Refs. 
\cite{AK,KMP93})~: 
\begin{eqnarray}
\omega_R &\simeq& v_F k s^{(0)} + O\,(T^4)~,          \label{omR} \\
\omega_I &\simeq& { (2F_0^\prime + 1) ((s^{(0)})^2 - 1) - (F_0^\prime)^2
                    \over
                    \tau F_0^\prime (F_0^\prime - (s^{(0)})^2 + 1)   }
                    + O\,(T^6)~,
                                                      \label{omI}
\end{eqnarray}
where $s^{(0)}$ is the root of collisionless dispersion relation
\begin{equation}
1 + F_0^\prime \chi(s^{(0)}) = 0~.                \label{drel1}
\end{equation}
The Landau parameter $F_0^\prime$ in Eqs.(\ref{omI}), (\ref{drel1})
is taken at $T=0$. The simple expression Eq.(\ref{omI}) for imaginary part 
of the frequency $\omega$ (long-dashed line on Fig. 3) reproduces 
the results of 
a numerical solution of the "exact" dispersion relation (\ref{drel}) 
(solid line on the same Fig. 3) with a good accuracy for temperatures 
$T < 2$ MeV. At larger temperatures, a slight increase of the damping due 
to temperature smearing of the Fermi distribution is obtained with
the dispersion relation Eq.(\ref{drel}). The difference between these
two solutions is of the order of the Landau damping rate in the pure 
mean field approach \cite{BV,BVA}. We see that thermal Landau 
damping is small for the case of GDR, at variance with the case of 
the breathing 
mode \cite{KLD97}, since the isovector Landau parameter is larger
than the isoscalar one for nuclear effective interactions at normal density. 
As a consequence a weaker coupling 
between single particle 
and collective motion is expected for isovector vibrations 
\cite{Hofm92}. 

The dominance of the collisional contribution 
to the total damping rate of the GDR is just expressed by an approximate
relation (short-dashed line on Fig. 3)
\begin{equation}   
-\mbox{Im}(\omega) \simeq {1 \over \tau}~,         \label{colcont}
\end{equation}
which was used, for instance, in Ref. \cite{Ayik} to calculate widths 
of giant resonances. The main deviation from formula (\ref{colcont})
in the dispersion relation Eq.(\ref{drel}) is caused 
by the exclusion of the $l=0$ harmonic from the collision integral
in the r.h.s. of Eq.(\ref{LVe}), i.e. due to taking into account
the particle number conservation. This results in a smaller absolute 
value of the Im($\omega$) (see Fig. 3).

As already discussed, a source of uncertainty in our calculations is 
given by the choice
of the nucleon-nucleon cross sections. In Fig. 2 we present calculations
with in-medium reduced cross sections (solid lines) and with free-space 
cross sections (dashed lines) in comparison with experimental widths for 
nuclei $^{208}$Pb (a) and $^{120}$Sn (b). There is a quite good reproduction 
of the experimental trend independently on the choice of cross sections.

In conclusion, we have studied the isovector response of heated 
spin-isospin symmetric nuclear matter on the basis of the linearized
Landau-Vlasov equation with a collision integral including memory
effects in the relaxation time approximation. 
The contribution of the thermal Landau damping 
into the total damping rate of isovector vibrations is found to be very small.
Thus the relaxation of the volume isovector mode is caused mainly
by nucleon-nucleon collisions. Increasing temperature shifts
the centroid energy of the isovector strength function to smaller
values. The calculated width is proportional to $1/\tau$ in 
the temperature region $T < 5$ MeV studied in this letter, that 
corresponds to the collisional damping of the isovector zero sound mode.
This leads to a $T^2$ behaviour of the $GDR-FWHM$ in the region
where the collective mode can propagate.

We have shown that some general GDR properties 
at high excitation energy can be obtained directly from Fermi liquid
theory. However the aim of this work is not to get a perfect agreement
with data for finite nuclei particularly at low temperatures. 
Indeed we are well aware that other contributions to the damping are 
missing in the present approach:
(i) The fragmentation width, observed in RPA calculations, which
can be interpreted as a Landau damping mechanism in finite
Fermi systems present also at zero temperature \cite{SB75}.
(ii) Thermal shape fluctuations \cite{OBB96,OBBB97}.
(iii) Fluctuations due to nucleon-nucleon correlations \cite{MD96}.


A recent statistical model analysis of $\gamma$-spectra produced by 
inelastic $\alpha$-scattering on $^{120}$Sn \cite{Gervais} resulted
in conclusion that neither the thermal fluctuation model of Ref.
\cite{OBBB97} nor the collisional damping model of this work could
reproduce data in details. We believe that the combination of 
the two models would give a much better agreement. 

Extension to isospin asymmetric nuclear systems, more suitable
for the $Pb$ case, can be performed following the approach
of ref.s \cite{Hans,CODI,MWF98}. We do not expect to have substantial
variations in the temperature behaviour of the isovector and isoscalar
modes unless very high charge asymmetries are reached \cite{Hans,CODI}.
However in charge asymmetric nuclei, new soft mode, different from
isovector and isoscalar ones, appears \cite{MWF98} due to collisional
coupling of proton and neutron vibrations, that requires further 
investigations.

Finally a comment on the validity of the formalism applied here
to the study of hot $GDR$ in nuclei to a broader context of
physical problems. Indeed it can be easily generalized to
the description of relative vibrations of any two-component
Fermi liquid with a mutual attraction, for instance of a
Coulomb plasma consisting of opposite charged fermions.
Another application could be to the oscillations of the
electronic cloud in metallic clusters, where momentum nonconserving
$l=1$ term in collision integral appears due to scattering of
electrons on impurities \cite{Mer70}.

\section*{ Acknowledgements }

We are gratefull to V.A. Plujko for help on derivation of relaxation 
time and for fruitfull discussions and to G.Q. Li for providing us with
data files containing in-medium nucleon-nucleon scattering cross sections.
Stimulating discussions with A. Bonasera and V.Baran are acknowledged.
One of us, A.B.L., acknowledges the warm hospitality and financial 
support of LNS-INFN.

\section*{ Appendix }

Here we will calculate the coefficient $\alpha^{(-)}$ in Eq. (\ref{tau})
for the collisional relaxation time $\tau$. For simplicity, we will use in 
derivation the Boltzmann-Uehling-Uehlenbeck (BUU) collision integrals without 
memory effects. However, as it is shown in Ref. (\cite{AYGS}), the collision 
integrals with memory effects give the result which can be obtained using 
Landau prescription $\tau = \tau_{BUU}/( 1 + (\hbar\omega_R/ 2 \pi T)^2 )$,
where $\tau_{BUU} = \hbar\alpha^{(-)}/T^2$ is the relaxation time given by
BUU collision integrals.   

Time evolution of the space-uniform isovector d.f. $f = f_n - f_p$
satisfies the equation (\cite{Ayik,KPS96})~:
\begin{equation}
{ \partial f({\bf p},t) \over \partial t } 
= I = I_{nn} + I_{np} - I_{pp} - I_{pn}~,                 \label{A1}
\end{equation}
where $I_{q_1q_2}$ stands for the collision integral of particles of 
the sort $q_1=n,p$ with particles of the sort $q_2=n,p$. Explicitly~:
\begin{eqnarray}
I_{q_1q_2}({\bf p_1},t) & = &
  {4 \over (2\pi\hbar)^6} \int\, d{\bf p_2} d{\bf p_3} d{\bf p_4}
w_{q_1q_2}({\bf p_1},{\bf p_2};{\bf p_3},{\bf p_4})
\delta(\triangle\epsilon) \delta(\triangle{\bf p})          \nonumber \\
 & & Q(f_{q_1}({\bf p_1}),f_{q_2}({\bf p_2});
     f_{q_1}({\bf p_3}),f_{q_2}({\bf p_4}))~,               \label{A2}
\end{eqnarray}
where
\[
Q(f_1,f_2;f_3,f_4) \equiv 
( 1 - f_1 ) ( 1 - f_2 ) f_3 f_4 - f_1 f_2 ( 1 - f_3 ) ( 1 - f_4 )~;
\] 
$w_{q_1q_2}({\bf p_1},{\bf p_2};{\bf p_3},{\bf p_4})$ is the spin-averaged
probability of two-body collisions with initial momenta 
$({\bf p_1},{\bf p_2})$ and final momenta $({\bf p_3},{\bf p_4})$;
$\triangle\epsilon = \epsilon_1 + \epsilon_2 - \epsilon_3 - \epsilon_4$,
$\epsilon_i = p_i^2/2m,~~i=1,2,3,4$, 
$\triangle{\bf p} = {\bf p_1} + {\bf p_2} - {\bf p_3} - {\bf p_4}$.
Neglecting the dependence of $w_{q_1q_2}$ on the d.f., we can write down
perturbations of collision integrals (\ref{A2}) keeping terms of the first
order in $\delta f_q$~:
\begin{eqnarray}
\delta I_{q_1q_2}({\bf p_1},t) = 
{4 \over (2\pi\hbar)^6} \int\, d{\bf p_2} d{\bf p_3} d{\bf p_4}
w_{q_1q_2}({\bf p_1},{\bf p_2};{\bf p_3},{\bf p_4})
\delta(\triangle{\bf p}) & &                                \nonumber    \\
\{ \alpha^{(1)} \, \psi_{q_1}({\bf p_1}) +   
  \alpha^{(2)} \, \psi_{q_2}({\bf p_2}) +
  \alpha^{(3)} \, \psi_{q_1}({\bf p_3}) +
  \alpha^{(4)} \, \psi_{q_2}({\bf p_4})   \}~,   & &    \label{A3}
\end{eqnarray}
where 
\begin{eqnarray}
\psi_q({\bf p}) & \equiv & \delta f_q({\bf p})
\left( \partial f_{eq}(\epsilon) \over \partial \epsilon \right)^{-1}~,~~
~~(q=n,p)~,                                                    \label{A4} \\
\alpha^{(i)}  & \equiv &
{ \delta
Q(f_{eq}(\epsilon_1),f_{eq}(\epsilon_2);f_{eq}(\epsilon_3),f_{eq}(\epsilon_4))
  \over \delta f_{eq}(\epsilon_i)    }                       \nonumber  \\
 & & { \partial f_{eq}(\epsilon_i) \over \partial \epsilon_i }
\delta(\triangle\epsilon)~,~~~~i=1,2,3,4~.                     \label{A5}
\end{eqnarray}
In Eq. (\ref{A3}) the isospin-symmetric nuclear matter is considered that
results in the same equilibrium d.f. for neutrons and protons. 

For the perturbation of the collision integral $I$ of the r.h.s. of
Eq. (\ref{A1}), assuming isotopic invariance ($w_{pp}=w_{nn}$,
$w_{pn}=w_{np}$) we can write after simple algebra~:
\begin{eqnarray}
 & & \delta I =
\delta I_{nn} + \delta I_{np} - \delta I_{pp} - \delta I_{pn} = \nonumber \\
 & & {4 \over (2\pi\hbar)^6} \int\, d{\bf p_2} d{\bf p_3} d{\bf p_4}
\delta(\triangle{\bf p})
     \{~( w_{pp} + w_{np} )\,( \alpha^{(1)} \, \psi({\bf p_1}) +
     \alpha^{(3)} \, \psi({\bf p_3})   )~ + ~      \nonumber \\
 & &  ( w_{pp} - w_{np} )\,( \alpha^{(2)} \, \psi({\bf p_2}) +
                             \alpha^{(4)} \, \psi({\bf p_4})   )
~ \}~,                                                \label{A6}
\end{eqnarray}
where $\psi({\bf p}) = \psi_n({\bf p}) - \psi_p({\bf p})$.

The triple integral over momenta in (\ref{A6}) can be taken using 
Abrikosov-Khalatnikov transformation \cite{AK}, which is valid in the 
limit $ T \ll \epsilon_F $~:
\begin{equation}
\int\, d{\bf p_2} d{\bf p_3} d{\bf p_4} \delta(\triangle{\bf p}) = 
{ (m^*)^3 \over 2 } \int\limits_0^\pi\, d\,\Theta 
{ \sin \Theta \over \cos(\Theta/2) }  
 \int\limits_0^\pi\, d\,\phi \int\limits_0^{2\pi}\, d\,\phi_2
 \int\limits_0^\infty\, d\,\epsilon_2 d\,\epsilon_3 d\,\epsilon_4~,  
                                                          \label{A7} 
\end{equation}
where $\theta = \widehat{(\hat p_1,\hat p_2)}$ is the angle between
momenta of colliding particles ($\hat p_i \equiv {\bf p_i}/|{\bf p_i}|$,
$i = 1,2,3,4$); $\phi$ is the angle between planes given by momenta
of incoming and outcoming particles~:
\begin{equation}
\cos \phi = { [\hat p_1 \times \hat p_2]\cdot[\hat p_3 \times \hat p_4]
              \over
            |[\hat p_1 \times \hat p_2]| |[\hat p_3 \times \hat p_4]|   }~;
\end{equation}
$\phi_2$ is azimutal angle of ${\bf p_2}$ in the system with z-axis along
${\bf p_1}$.

We decompose the perturbation of the d.f. into spherical harmonics~:
\begin{equation}
\psi({\bf p}_i,t) = \sum\limits_{l,m}\, \alpha_{lm}(p_i,t)
                    Y_{lm}(\hat p_i)~,~~~i=1,2,3,4~,        \label{A8}
\end{equation}
where coefficients $\alpha_{lm}$ can be taken on the Fermi surface,
since $T \ll \epsilon_F$.

According to Refs. \cite{Ayik,KMP93,BS70}, the partial relaxation time
$\tau_l$ is defined as follows~:
\begin{eqnarray}
{ 1 \over \tau_l^{BUU} }& = &
- { \int\limits_0^\infty d\epsilon \int d\Omega_{\hat p} \,
    Y_{lm}^*(\hat p) \, \delta I({\bf p},t)    \over
    \int\limits_0^\infty d\epsilon \int d\Omega_{\hat p} \,
    Y_{lm}^*(\hat p) \, \delta f({\bf p},t) } =         \nonumber  \\
 & & { \int d\Omega_{\hat p} \, Y_{lm}^*(\hat p) \,
       \overline{\delta I}(\hat p,t) \over \alpha_{lm}  }~,   \label{A9}
\end{eqnarray}
where
\begin{equation}
\overline{\delta I}(\hat p) = \int\limits_0^\infty d\epsilon
\, \delta I({\bf p})~,~~~~~~~\epsilon = p^2/2m~.               \label{A10}
\end{equation}

Using (\ref{A7}), after somewhat lengthy but standard calculations,
we come to the expression~:
\begin{equation}
{ 1 \over \tau_l^{BUU} } = { (m^*)^3 T^2 \over 12 \pi^2 \hbar^6 }
\left\{ < w_{pp}\Phi_l^{(+)} > + 2< w_{np}\Phi_l^{(-)} > \right\}
\equiv {T^2 \over \kappa_l}~,
                                                             \label{A11}
\end{equation}
where angular brackets denote the averaging over angles $\theta$ and
$\phi$ \cite{KPS96,KMP93}~:
\begin{equation}
< F(\theta,\phi) >  \equiv {1 \over 2\pi} \int\limits_0^\pi d\theta
{ \sin\theta \over \cos{\theta \over 2} }
\int\limits_0^\pi d\phi \, F(\theta,\phi)~;                   \label{A12}
\end{equation}
\begin{eqnarray}
\Phi_l^{(+)} & = & 1 + P_l(\hat p_2 \cdot \hat p_1) -
                   P_l(\hat p_3 \cdot \hat p_1) -
                   P_l(\hat p_4 \cdot \hat p_1)~,             \label{A13} \\
\Phi_l^{(-)} & = & 1 - P_l(\hat p_2 \cdot \hat p_1) -
                   P_l(\hat p_3 \cdot \hat p_1) +
                   P_l(\hat p_4 \cdot \hat p_1)~.             \label{A14} 
\end{eqnarray}
A factor 2 at the second term in curly brackets of (\ref{A11}) is due to
half momentum space integration over $d{\bf p_3}$ in the l.h.s.
of Eq. (\ref{A7}) \cite{AK,SB70}. The arguments of the Legendre polynomials 
in functions (\ref{A13}),(\ref{A14}) are~:
\begin{eqnarray}
\hat p_2 \cdot \hat p_1 & = & \cos\theta~,                    \nonumber \\
\hat p_3 \cdot \hat p_1 & = & \cos^2{\theta \over 2} +
\sin^2{\theta \over 2} \cos\phi~,                             \nonumber \\
\hat p_4 \cdot \hat p_1 & = & \cos^2{\theta \over 2} -
\sin^2{\theta \over 2} \cos\phi~.                             \nonumber
\end{eqnarray}
For $l=1,2$ and $\infty$ we have~:
\begin{eqnarray}
 & & \Phi_1^{(+)} = 0~,~~~~~~\Phi_1^{(-)} = 4\sin^2{\theta \over 2}
                             \sin^2{\phi \over 2}~,           \label{A15} \\
 & & \Phi_2^{(+)} = 3\sin^4{\theta \over 2}\sin^2\phi~,~~~~~~
     \Phi_2^{(-)} = 3\sin^2\theta\sin^2{\phi \over 2}~,       \label{A16} \\
 & & \Phi_\infty^{(+)} = \Phi_\infty^{(-)} = 1~.                \label{A17}
\end{eqnarray}

Collision probabilities can be expressed in terms of cross sections
as follows~:
\begin{eqnarray}
w_{pp} & = & { (2\pi\hbar)^3 \over 2 \mu^2 } 
             { d\sigma_{pp} \over d\Omega_{c.m.} }~,         \label{A18} \\
w_{pn} & = & { (2\pi\hbar)^3 \over 2 \mu^2 } 
             { d\sigma_{pn} \over d\Omega_{c.m.} }~,         \label{A19} 
\end{eqnarray}
where $d\sigma_{pp}/d\Omega_{c.m.}$ and $d\sigma_{pn}/d\Omega_{c.m.}$ are
differential cross sections of pp and np scattering; 
$d\Omega_{c.m.} = \sin\theta_{c.m.} d\theta_{c.m.} d\phi_{c.m.}$;
$\theta_{c.m.}$ and $\phi_{c.m.}$ are polar and azimutal scattering angles
in the center of mass system of colliding particles; $\mu = m/2$ is
the reduced mass. Differential cross sections $d\sigma_{pp}/d\Omega_{c.m.}$
and $d\sigma_{pn}/d\Omega_{c.m.}$ depend on the relative momentum 
$p\prime = |{\bf p_1} - {\bf p_2}|/2$ of scattered particles and on the
polar angle 
\[
\theta_{c.m.} = \arccos\left( { ({\bf p_1} - {\bf p_2}) \cdot
                                ({\bf p_3} - {\bf p_4})       \over
                                |{\bf p_1} - {\bf p_2}| 
                                |{\bf p_3} - {\bf p_4}|           
                        } \right)~.
\]
For particles scattered on the Fermi surface, we have~:
\[
p\prime = p_F \sin{\theta \over 2}~,~~~~~\theta_{c.m.} = \phi~.  
\]
In a particular case of isotropic energy independent cross sections the result
of Ref. \cite{Ayik} is recovered~:
\begin{eqnarray}
{1 \over \tau_1^{BUU}} & = & {32 \over 9} {m \sigma_v \over \hbar^3} T^2~,
                                                      \label{A20} \\
{1 \over \tau_2^{BUU}} & = & {32 \over 15} {m \sigma_s \over \hbar^3} T^2~,
                                                      \label{A21}
\end{eqnarray}
where $\sigma_v = \sigma_{np}/2$, 
$\sigma_s = (\sigma_{nn} + \sigma_{pp} + 2\sigma_{np})/4$,
$\sigma_{np} = (4\pi) d\sigma_{np}/d\Omega_{c.m.} \simeq 50$ mb,
$\sigma_{nn} \simeq \sigma_{pp} = (2\pi) d\sigma_{pp}/d\Omega_{c.m.} 
\simeq 25$ mb.

We derived relaxation times $\tau_1^{BUU}$, $\tau_2^{BUU}$ and 
$\tau_\infty^{BUU}$ using pp and nn energy and angular dependent 
differential cross sections calculated with Bonn A potential 
\cite{LiMach,Li}. Results of these calculations both with vacuum and 
in-medium cross sections at normal nuclear density are given in the Table. 
It is seen from the Table that always 
$\tau_1^{BUU} \simeq \tau_2^{BUU} \simeq \tau_\infty^{BUU}$. This gives 
an idea to put the same value for all relaxation times $\tau_l^{BUU}$,
i.e. to apply usual relaxation time approximation. Thus, we obtain
the collision integral of Eq. (\ref{LVe}) with relaxation time $\tau$
given by Eq. (\ref{tau}), where
\[
\alpha^{(-)} = \left[ {\hbar \over 3} 
                      (\kappa_1^{-1} + \kappa_2^{-1} + \kappa_\infty^{-1})
               \right]^{-1} = 2.3~(5.4)~\mbox{MeV}
\]
for vacuum (in-medium) cross sections of Ref. \cite{Li}.

\newpage

\begin{description}
\item[Table] Parameters $\kappa_l,~l=1,2$ and $\infty$ (MeV$^2$ fm/c) 
defined in Eq. (\ref{A11}) at various choises of 
nucleon-nucleon scattering cross sections.
\end{description}
\begin{tabular}{|c|c|c|c|}
\hline
NN cross sections         & $\kappa_1$ & $\kappa_2$ & $\kappa_\infty$ \\
\hline
Vacuum of Ref. \cite{Li}  &   503      & 491        & 401     \\
\hline
In-medium at              &   1123     & 1068       & 1003    \\
$\rho=\rho_0$             &            &            &         \\  
of Ref. \cite{Li}         &            &            &         \\
\hline
Isotropic energy-         &   920      & 1022       & 818     \\
independent               &            &            &         \\
of Ref. \cite{Ayik}       &            &            &         \\
\hline
\end{tabular}

\newpage

\section*{Figure captions}

\begin{description}

\item[Fig. 1] Photoabsorption cross section by an excited nucleus
$^{208}$Pb as a function of photon energy for temperatures $T=0$ MeV 
(solid line), $T=2$ MeV (short-dashed line) and $T=4$ MeV
(long-dashed line) calculated with in-medium cross sections of 
Ref. \cite{Li}.

\item[Fig. 2] Centroid energy $E_{GDR}$ (MeV) and $FWHM$ (MeV) for 
the $GDR$ mode in nuclei $^{208}$Pb (a) and $^{120}$Sn (b) as functions 
of temperature calculated with in-medium cross sections (solid lines)
and with free space cross sections (dashed lines). Points with 
errorbars show the experimental widths from Ref. \cite{Ram1} -- $^{208}$Pb
and from Ref. \cite{Ram2} -- $^{120}$Sn.   

\item[Fig. 3] Temperature dependence of real and imaginary parts of 
the pole of the response function (\ref{chicoll}) 
-- upper and lower solid lines respectively, and 
collisional width $\Gamma_{coll}=2\hbar/\tau$ -- short-dashed line. 
The long-dashed line shows the imaginary part of pole as given by 
the approximate  Eq.(\ref{omI}). All values are in MeV. 
In-medium cross sections are used.

\end{description}


\begin{thebibliography}{99}
\bibitem{Comm1} To avoid misunderstanding, we stress that Landau damping
in finite nuclei at T=0, which produces the fragmentation width, is not 
discussed in this work.
\end{thebibliography}
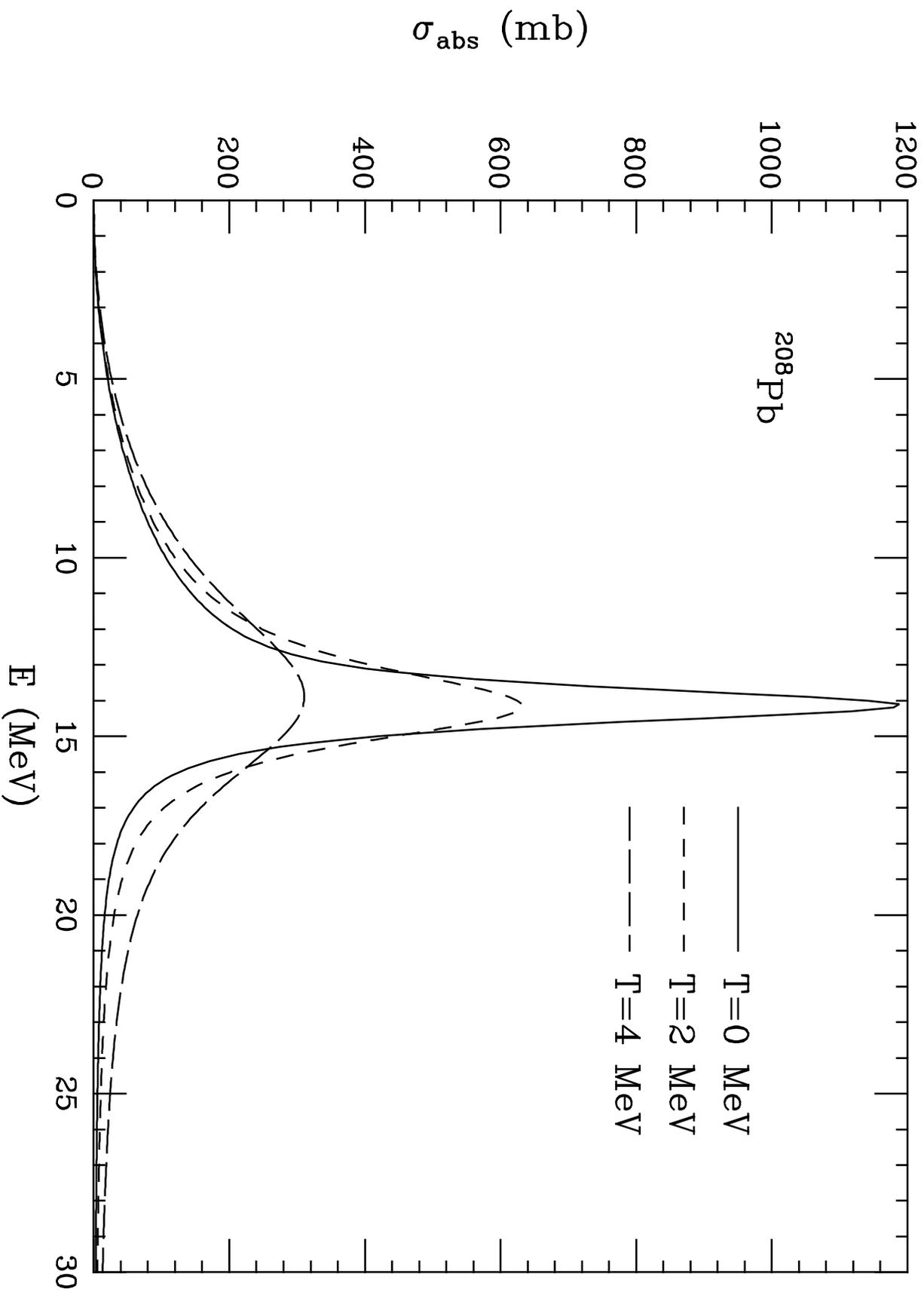
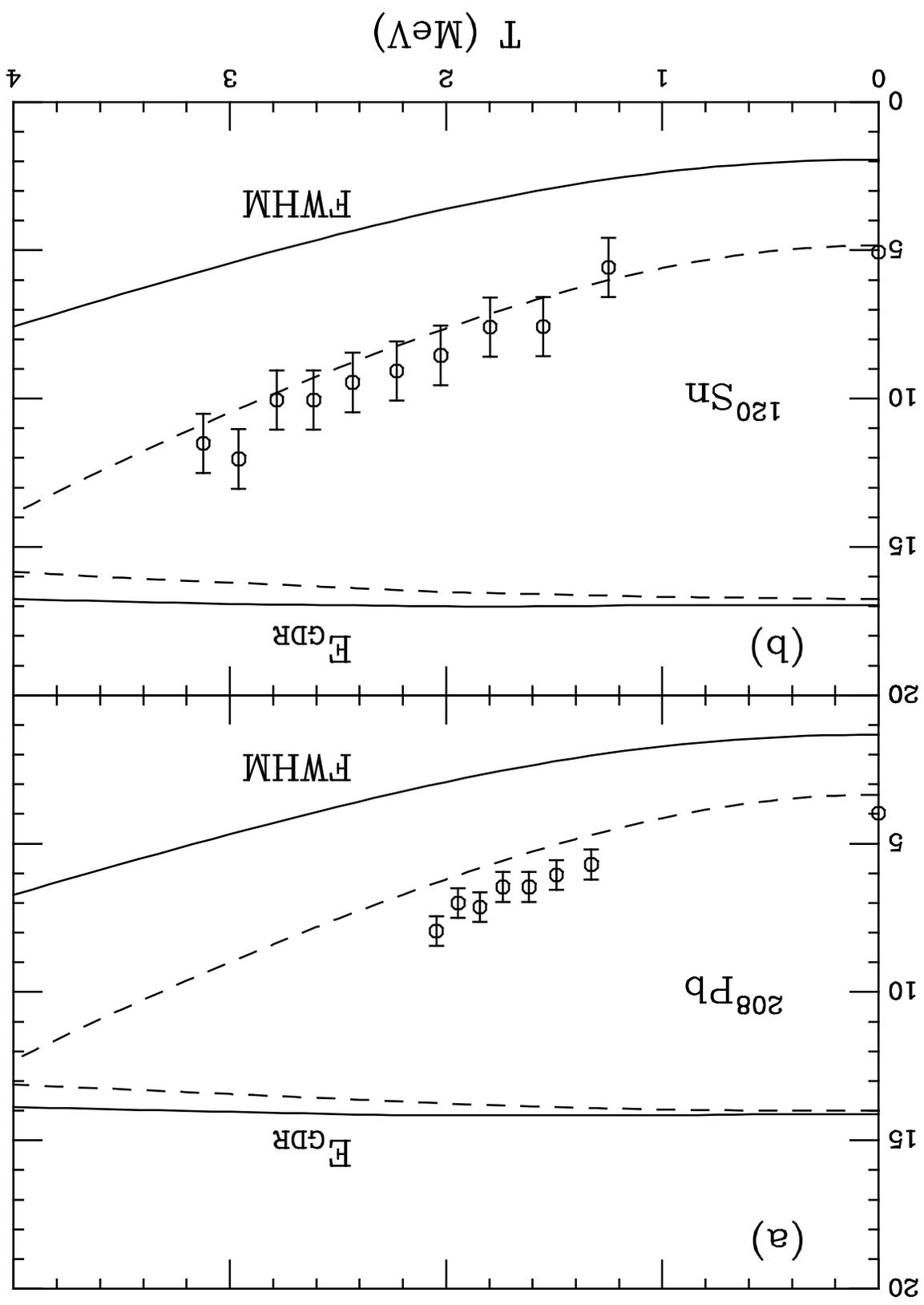
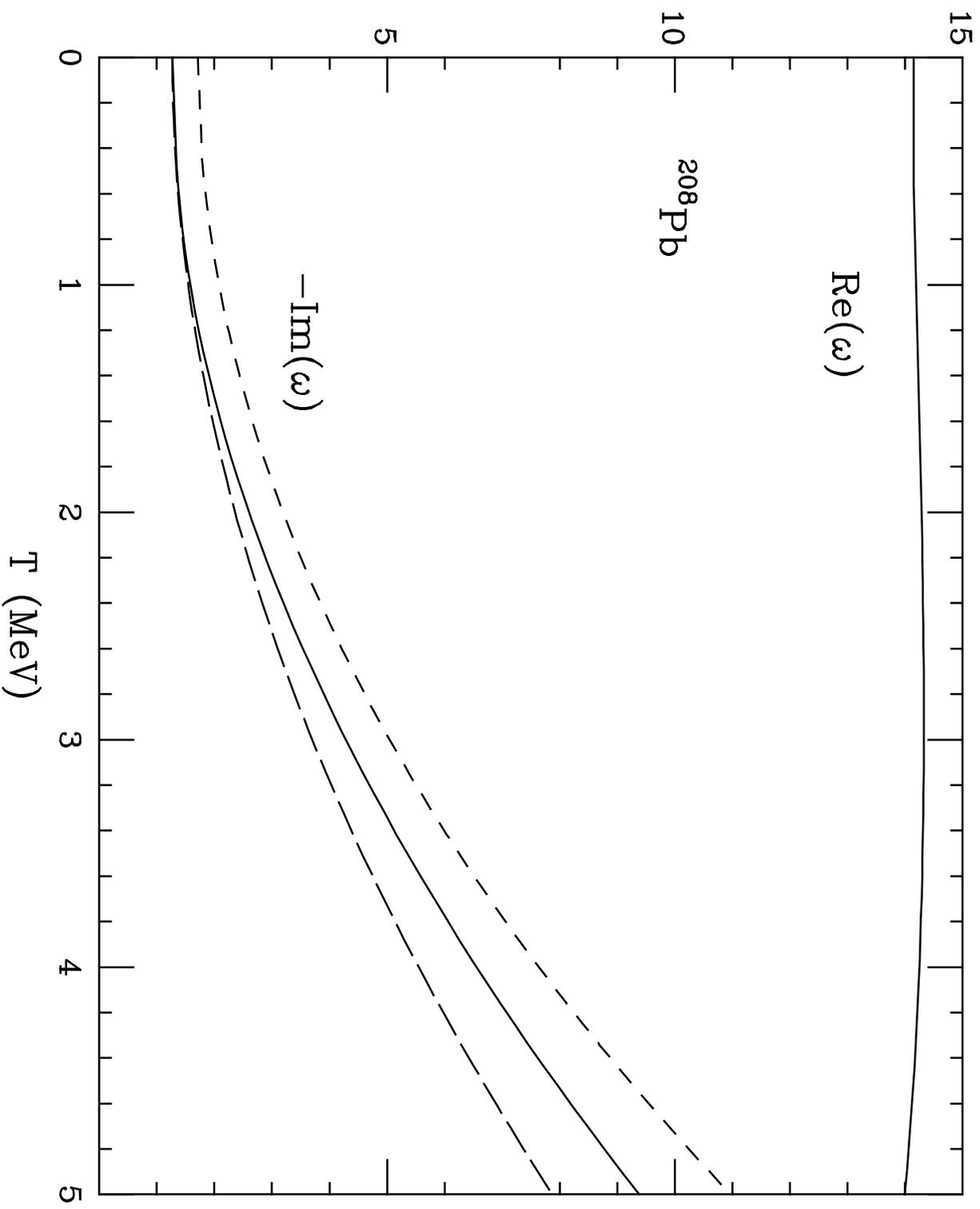
\end{document}